\documentclass{article}
\usepackage{graphicx}

\def\half{\frac{1}{2}}
\def\nn{\nonumber}
\def\df{\partial}

\begin{document}

\title{Tachyonic models of dark matter}
\author{Igor Nikitin\\
Department of High Performance Analytics\\
Fraunhofer Institute for Algorithms and Scientific Computing\\
Schloss Birlinghoven, 53757 Sankt Augustin, Germany\\
\\
igor.nikitin@scai.fraunhofer.de}
\date{}

\maketitle

\begin{abstract}
We consider a spherically symmetric stationary problem in General Relativity,
including a black hole, inflow of normal and tachyonic matter and outflow
of tachyonic matter. Computations in a weak field limit show that the 
resulting concentration of matter around the black hole 
leads to gravitational effects equivalent to those associated 
with dark matter halo. In particular, the model reproduces 
asymptotically constant galactic rotation curves,
if the tachyonic flows of the central supermassive black hole
in the galaxy are considered as a main contribution.
\end{abstract}

\begin{figure}
\centering
\includegraphics[width=\textwidth]{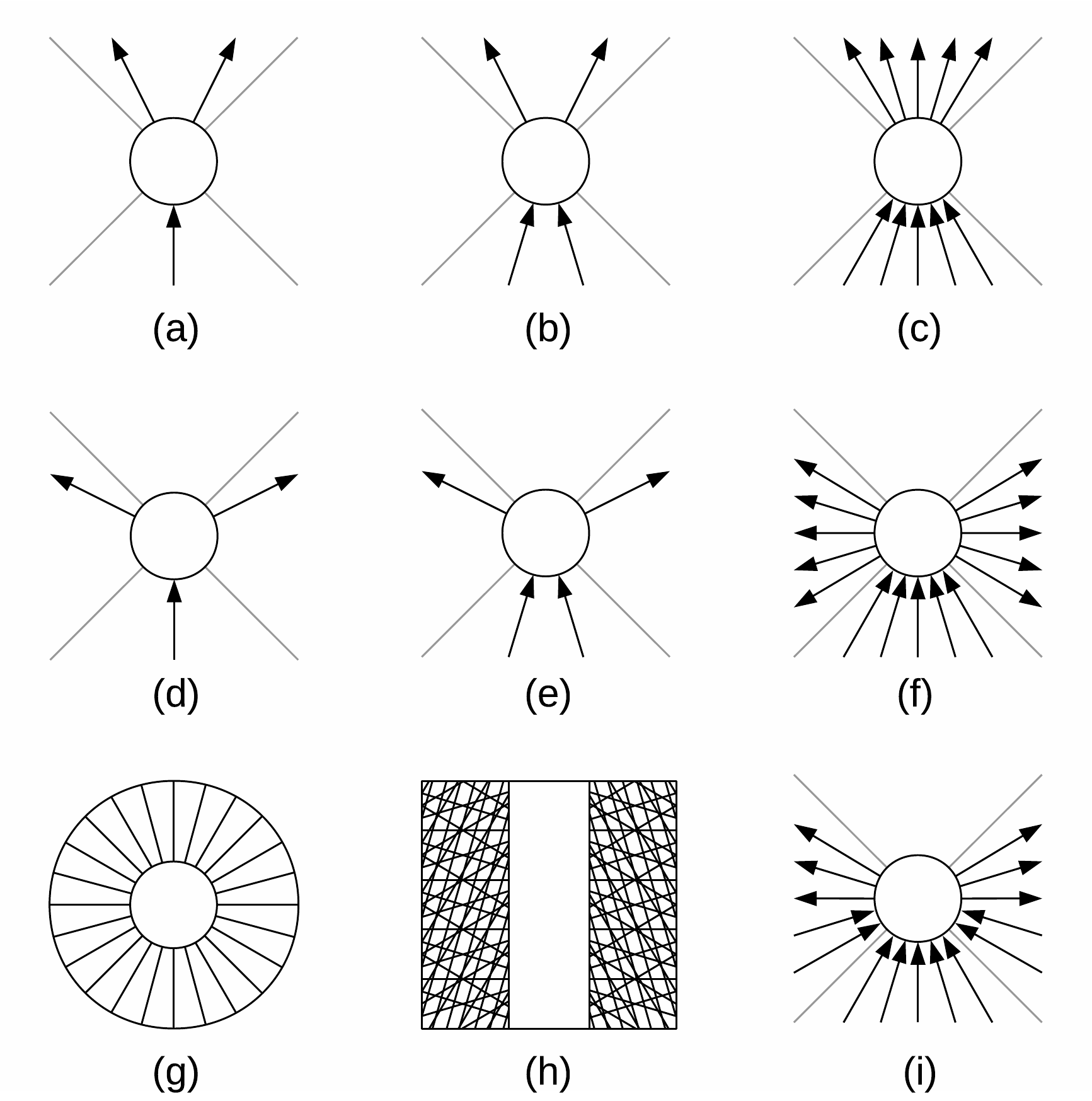}
\caption{Elementary processes (see description in text).}
\label{f1}
\end{figure}

\section{Introduction}\label{intro}
Tachyonic models for the description of dark matter have appeared recently.
Papers \cite{shiu,Frolov,Bagla} consider a model of
tachyonic scalar field, based on the action
\begin{eqnarray}
&&S[T(x)]=
-\int d^{4}x\sqrt{-g}\ V(T)\sqrt{1+\nabla_{\mu}T\nabla^{\mu}T},\label{ST}
\end{eqnarray}
where $T(x)$ is a scalar field, $V(T)$ is a given potential function, 
$\nabla_{\mu}$ is covariant derivative and $d^{4}x\sqrt{-g}$
is invariant integration measure of General Relativity.
In this model the distribution of energy-momentum can be decomposed
to a sum of pressureless liquid, interpreted as dark matter,
and a negative pressure medium, interpreted as dark energy.
After fine tuning of the parameters, the model becomes compatible
with the standard Big Bang cosmology, providing an explanation
for the observed accelerated expansion of the universe.

The paper \cite{Davis} takes a different approach, considering geodesic flows 
of particles, described by the action
\begin{eqnarray}
&&S_{\pm}[x(\tau)]=\mp m\int d\tau\sqrt{\mp\dot x_{\mu}\dot x^{\mu}},\label{PNT}
\end{eqnarray}
where $x(\tau)$ is the world line of the particle,
$\dot x=dx/d\tau$, the upper sign corresponds to normal matter,
the lower sign to tachyonic matter and the metric signature 
$(-,+,+,+)$ is chosen. Considering such flows in non-stationary
FRW universe, the paper shows that the curved metric forces 
the tachyonic world lines to turn back in time, leading 
to self-annihilation of the tachyons and their disappearance from the universe.
We note, however, that a reserve of the tachyons can be renewed,
if one admits that the tachyons are created in local singularities,
under event horizons, and escape from the black holes 
along spacelike world lines. In this paper we will consider
а tachyonic model in the context \cite{Davis}, as a geodesic flow of particles,
and focus our attention on the role that can be played by black holes 
in this model. Particularly, we study the following scenario.

Let's consider an isolated spherically symmetric black hole and a flow 
of particles of normal matter 
isotropically falling into it from infinity. From a point
of view of a distant observer the particles slow down at event horizon and never
intersect it, while in a coordinate system moving together with 
the particles they pass through event horizon and move further 
towards singularity. The black hole acts like a natural accelerator,
where the particles are boosted to extremely high energies. 
This opens an opportunity for new physics, in particular, 
we will admit that high energetic collisions of particles 
lead to generation of tachyons.

Kinematically the processes of transformation of normal matter to tachyons
are allowed. The tachyons are superluminal particles possessing opposite 
sign in mass-shell condition and propagating along spacelike world lines,
outside the light cone. Fig.\ref{f1} shows a collection of 
various processes happening with normal particles and tachyons. 
Here time axis is vertical, 
space axis is horizontal and light cones are shown by grey lines 
(except of fig.\ref{f1}g, showing purely spatial projection).
The vector of energy-momentum of every particle is directed along 
its world line. 

Fig.\ref{f1}a shows a process of decay of one normal
particle to two normal particles: $(m,0)\to(m/2,\vec p)+(m/2,-\vec p)$
with $(m/2)^{2}-\vec p^{\;2}>0$. Fig.\ref{f1}d shows a decay of the same normal
particle to two tachyons: $(m/2)^{2}-\vec p^{\;2}<0$. In both case 
the conservation of energy-momentum is satisfied. 

Fig.\ref{f1}b shows a collision
of two normal particles leading to creation of two other normal particles
$(E,\vec p_{1})+(E,-\vec p_{1})\to(E,\vec p_{2})+(E,-\vec p_{2})$,
with $E^{2}-\vec p_{1}^{\;2}>0$ and $E^{2}-\vec p_{2}^{\;2}>0$.
Fig.\ref{f1}e shows creation of two tachyons, with $E^{2}-\vec p_{2}^{\;2}<0$.
Here the conservation of energy-momentum is satisfied as well. 

Therefore the question here is not in kinematic feasibility
but in the existence of interaction vertices
allowing these transformations. In this paper we assume that interaction
vertices for transformation of normal particles to tachyons exist 
and are activated at high energies, achievable only under event horizons.
At low energies the vertices are suppressed. 

A possible mechanism
of this suppression can be an existence of a supermassive normal particle
to which the tachyons are directly coupled, so that creation of 
freely propagating tachyons requires an overcoming of a high mass barrier. 
The other mechanism can be direct dependence of vertex function on the energy.
Here we will not fix this mechanism and just assume that
under event horizon the normal matter will be converted to tachyons 
before falling into the singularity. 

Since the tachyons 
are superluminal particles, they are not confined in the black hole and 
can leave it along spacelike world lines. In this way the falling matter 
is returned
back to our side of the universe in the form of outgoing flow of tachyons.
Outside the black hole the tachyons do not interact directly 
with the normal matter and with each other and move freely along 
the geodesics. On the other hand, since the
tachyonic flows have non-zero density of energy-momentum,
they are able to curve the space-time and produce observable 
gravitational effects. Thus the tachyonic flows outside of
event horizons behave like invisible type of matter 
interacting with the normal matter only gravitationally,
making it a suitable candidate for the role of astrophysical dark matter.

In further detail, Fig.\ref{f1}c shows a process of multiple collision
of normal particles transformed to a shower of normal particles.
Kinematically the outgoing particles can occupy the future light cone
of collision point. Fig.\ref{f1}f shows analogous process producing
a shower of tachyons, occupying the exterior part of the light cone. 
Conservation of momentum can be easily fulfilled e.g. by considering 
spherically symmetrical incoming and outgoing flows.
Conservation of energy is recorded as equality of total energy 
for incoming and outgoing flows. It is a single integral equation 
leaving enough degrees of freedom for setting detailed distributions
of energy in the flows. 

For the case of tachyons, there are
world lines in the exterior of the light cone going forward in time
and there are world lines going backward in time. These types of world lines
cannot be separated in Lorentz invariant way, i.e. this separation
depends on the choice of a reference frame. The vector of energy-momentum
is directed along the world lines, so that world lines going backward in time
possess negative energy. They can be also considered as the world lines
going forward in time and possessing positive energy, like shown on 
Fig.\ref{f1}i. 

All physically meaningful relativistic models
are invariant under reversal of the direction of the world lines,
which actually just a convention where the world line starts and
where ends. E.g. the action of tachyons is the length of the world line, 
invariant under its reversal. 
We will see further that tensor of energy-momentum 
for tachyonic flow is quadratic in velocities and reversal of the velocities 
does not change it either. Thus the reversal of the world lines changes only 
the interpretation, while the flows depicted 
on Fig.\ref{f1}f and Fig.\ref{f1}i
produce physically equivalent answers. In summary, Fig.\ref{f1}i
shows incoming flow of normal matter, incoming flow of tachyons
and outgoing flow of tachyons, while outgoing flow of normal
matter is completely blocked by the black hole.
Note that in this interpretation all flows have positive energy.

Further, we will consider not only a single collision event,
but a stationary process supported by these flows.
I.e. we consider incoming flows of normal matter and tachyons
as permanently sourced at infinity and absorbed by the black hole
and outgoing flow of tachyons as permanently emitted by
the black hole and sinked at infinity. 
Although such stationarity is not absolutely necessary,
it simplifies a lot the computation of gravitational effects.
To make the process stationary, we need to superimpose 
multiple copies of Fig.\ref{f1}i shifted along time axis and 
form the distribution of world lines shown on Fig.\ref{f1}h. 

Also for our convenience, we will consider spherically symmetric 
distributions and
restrict computations to a finite spherical layer $r\in[r_{1},r_{2}]$, 
schematically shown on spatial projection Fig.\ref{f1}g.
In this way all unknown processes become located outside of 
the considered domain: at $r<r_{1}$ there are processes of transformation 
of normal matter to tachyons, at $r>r_{2}$ 
there are processes permanently sourcing incoming flows and
sinking outgoing flows. 
Inside spherical layer there are only geodesic flows of matter and 
curved space-time, whose known physics allows us to perform 
straightforward computations.

In Section \ref{sec_tachyo} we consider dynamics of tachyons in comparison 
with the dynamics of normal particles. In Section \ref{sec_set} we set
energy-momentum tensor for the above described spherically 
symmetric stationary problem. In Section \ref{sec_efe} we solve Einstein
equations in the limit of weak fields. In Section \ref{sec_disc}
we discuss the obtained results and outline possible extensions of the model.

\section{Dynamics of tachyons}\label{sec_tachyo}

The world lines of the particles are stationary points of the action
\begin{eqnarray}
&&S_{\pm}[x(\tau)]=\mp m\int d\tau\sqrt{\mp g_{\mu\nu}\dot x^{\mu}\dot x^{\nu}},\label{PNT2}
\end{eqnarray}
We remind that General Relativity (GR) 
distinguishes between upper tensor indices 
(called contravariant) and lower tensor indices (called covariant)
and
\begin{itemize}
\item $g^{\mu\nu}$ is inverse to $g_{\mu\nu}$,
\item metric tensor is used to raise and lower the indices,\\
e.g. $x_{\mu}=g_{\mu\nu}x^{\nu}$, $x^{\mu}=g^{\mu\nu}x_{\nu}$, 
\item summation over repeating indices is everywhere assumed,
\item length element in space-time is $ds^{2}=g_{\mu\nu}dx^{\mu}dx^{\nu}$,
\item invariant integration measure is $d^{4}x\sqrt{-g}$, 
where $g=\det g_{\mu\nu}$,
\item $\nabla$ denotes covariant derivative, $\Gamma$
are Christoffel symbols:
\begin{eqnarray}
&&\Gamma^{\mu}_{\nu\lambda}=\half g^{\mu\rho}(\df_{\lambda}g_{\rho\nu}
+\df_{\nu}g_{\rho\lambda}-\df_{\rho}g_{\nu\lambda})\nn\\
&&\nabla_{\alpha}V^{\mu}=\df_{\alpha}V^{\mu}+\Gamma^{\mu}_{\alpha\lambda}V^{\lambda},\nn\\
&&\nabla_{\alpha}T^{\mu\nu}=\df_{\alpha}T^{\mu\nu}+\Gamma^{\mu}_{\alpha\lambda}T^{\lambda\nu}+\Gamma^{\nu}_{\alpha\lambda}T^{\mu\lambda},\ \mbox{etc.}\nn
\end{eqnarray}
\end{itemize}

The integral (\ref{PNT2}) defines total length of the world line
in curved metric and its extremum corresponds to geodesics.
Special relativity (SR) corresponds to flat metric
$\eta_{\mu\nu}=\mbox{diag}(-1,1,1,1)$ and straight geodesic lines.

Upper sign in the action corresponds to timelike world lines $ds^{2}<0$,
i.e. the particles of normal matter (in the literature also called 
tardyons or bradyons). Lower sign  corresponds to spacelike world lines 
$ds^{2}>0$, the tachyons. Overall sign is selected in a way that
canonical momentum $p_{\mu}=\delta S_{\pm}/\delta\dot x^{\mu}$
in contravariant recording
\begin{eqnarray}
&&p^{\mu}=m\dot x^{\mu}/\sqrt{\mp\dot x^{\alpha}\dot x_{\alpha}}\label{pdef}
\end{eqnarray}
would have positive temporal component, for the world lines directed 
in the future and $m>0$. This convention ensures 
positive energy for the particles. Mass shell condition has a form
\begin{eqnarray}
&&p^{\mu}p_{\mu}=\mp m^{2},\label{mass2}
\end{eqnarray}
so that for normal particles $m$ can be identified with the mass 
of the particle. Tachyons are often described as particles with
imaginary mass, but we will consider $m$ as real parameter
and for tachyons explicitly fix a different sign in the mass shell condition.

\vspace{2mm}\noindent
{\it Remark about negative masses:} 
the case of $m<0$ is usually called exotic matter
and corresponds indeed to very unusual effects, like repelling
gravitational force (anti-gravitation). For tachyons the case $m<0$
would be double exotic, describing spacelike world lines
with momentum vector opposite to the direction of the world line.
In our model we will use only positive masses.
Although negative masses are theoretically possible, they are not needed
for a moment.

\vspace{2mm}\noindent
{\it Remark about causality principle:} 
involving tachyons in the model, one could expect causality violations, since one can make tachyons to propagate back in time simply by a change of coordinate frame.  However, we have seen that the reversal of the world line of the tachyon leaves its physics invariant. Also, a possibility to transmit information by tachyons implies an ability to interact with them, while in our model all points of direct interaction are hidden under event horizons. Although the tachyons can interact with the normal matter gravitationally, these effects, as all effects related to the dark matter, are supposedly detectable only on large astronomical scale. We are curious if it will be possible to construct a measurable  violation of causality principle under these conditions. In this relation we refer to classical work of Wheeler and Feynman \cite{WF1,WF2} about advanced and retarded interactions, where the questions of causality violation have been analyzed in detail. 

\section{Setting energy-momentum tensor}\label{sec_set}

Tensor of energy-momentum is defined by the formula
\begin{eqnarray}
&&T^{\mu\nu}(x)=2(-g)^{-1/2}\;\delta S/\delta g_{\mu\nu}(x)\nn
\end{eqnarray}
and for a pointlike particle can be written as
\begin{eqnarray}
&&T^{\mu\nu}=(-g)^{-1/2}\;m\int d\tau\;\delta(x(\tau)-x)\;\dot x^{\mu}\dot x^{\nu}/\sqrt{\mp\dot x^{\alpha}\dot x_{\alpha}},\nn
\end{eqnarray}
or equivalently:
\begin{eqnarray}
&&T^{\mu\nu}=\rho u^{\mu}u^{\nu},\ \rho=(-g)^{-1/2}\;m\int ds\;\delta(x(s)-x),\
u^{\mu}=(dx^{\mu}(s)/ds)|_{x(s)=x}.\nn
\end{eqnarray}
Here $ds=(|ds^{2}|)^{1/2}$ introduces natural parametrization on the world line,
$u^{\mu}$ is a tangent vector to the world line, with proper normalization:
\begin{eqnarray}
&&u^{\mu}u_{\mu}=\mp1,\ u^{\mu}=p^{\mu}/m.\nn
\end{eqnarray}
The factor $(-g)^{-1/2}$ makes $\rho$ invariant (scalar) under diffemorphisms 
of $x$ and corresponding transformation of metric. $\rho(-g)^{1/2}d^{4}x$ 
gives a mass element, $\rho$ represents a density of mass 
per invariant volume $(-g)^{1/2}d^{4}x$, while $\rho(-g)^{1/2}$ represents 
a density of mass per standard volume $d^{4}x$. In the considered case
the function $\rho(-g)^{1/2}$ is singular, it describes a positive mass 
localized on the world line, uniformly distributed on it with respect 
to the natural parameter. The shift of points along the world line
$x(s)\to x(s+ds)$ 
preserves this mass distribution. Foliating the space-time to such world lines, 
we have a tensor of energy-momentum for the flow of particles in the form:
\begin{eqnarray}
&&T^{\mu\nu}=\rho u^{\mu}u^{\nu},\ \rho>0,\label{dust}
\end{eqnarray}
where $u^{\mu}$ is the velocity of the flow in the given point. 
The density $\rho(-g)^{1/2}$ is again invariant under the shifts of points 
along the world lines, i.e. is preserved by the flow, 
following standard continuity equation
$\df_{\mu}(\rho(-g)^{1/2}u^{\mu})=0$. Using the identity 
$\df_{\mu}((-g)^{1/2}V^{\mu})=(-g)^{1/2}\nabla_{\mu}V^{\mu}$ 
from \cite{Dirac}, we can rewrite this equation in covariant form as
\begin{eqnarray}
&&\ \nabla_{\mu}(\rho u^{\mu})=0.\label{nablarho}
\end{eqnarray}
Note that (\ref{dust}) and (\ref{nablarho}) are
well known formulae for a pressureless liquid or for a dust,
we just ensure that their derivation does not rely upon normal or tachyonic
type of matter, so they are also valid for tachyonic flows.

Further in this section we will use flat metric,
fix spherical coordinates $x=(t,r,\theta,\phi)$
and consider the flows depicted on Fig.\ref{f1}f. 
The world lines can be parametrized as follows:
\begin{eqnarray}
&&x^{\mu}_{+}(s;\beta,t_{0},\theta_{0},\phi_{0})
=(s\cosh\beta+t_{0},s\sinh\beta,\theta_{0},\phi_{0}),\ 
\mbox{for normal matter;}\nn\\
&&x^{\mu}_{-}(s;\beta,t_{0},\theta_{0},\phi_{0})
=(s\sinh\beta+t_{0},s\cosh\beta,\theta_{0},\phi_{0}),\
\mbox{for tachyons.}\nn
\end{eqnarray}
The velocities are
\begin{eqnarray}
&&u^{\mu}_{+}(\beta)=(\cosh\beta,\sinh\beta,0,0),\ 
\mbox{for normal matter;}\nn\\
&&u^{\mu}_{-}(\beta)=(\sinh\beta,\cosh\beta,0,0),\
\mbox{for tachyons.}\nn
\end{eqnarray}
The density function satisfying mass conservation (\ref{nablarho})
has a form $\rho(r)=r^{-2}\rho_{1}$ with a constant $\rho_{1}>0$.
Such dependence is clear from geometrical point of view:
the density of the world lines increases towards the origin
inverse quadratically with the distance. It is also clear physically:
considering particles in a shell $[r,r+dr]$ moving at a constant
speed towards the origin, the mass density will have the same
behavior. The overall flow distribution can be parametrized
as follows:
\begin{eqnarray}
&&T^{\mu\nu}(r)=r^{-2}\int d\beta(\rho_{+}u^{\mu}_{+}u^{\nu}_{+}
+\rho_{-}u^{\mu}_{-}u^{\nu}_{-}),\nn
\end{eqnarray}
where $\rho_{\pm}(\beta)>0$ are arbitrary profile functions.
Here $\beta<0$ corresponds to inflow, $\beta>0$ to outflow.
Since the flow of normal matter depicted on Fig.\ref{f1}f has
no outflow component, one can formally extend
$\rho_{\pm}(\beta)\geq0$ to the whole axis and set
$\rho_{+}(\beta)=0$ at $\beta>0$.

We will also require that the flows are energetically balanced,
i.e. the energies of incoming and outgoing flows coincide.
This is equivalent to vanishing total flow of energy through
the spatial 2-spheres, i.e. $T^{tr}=T^{rt}=0$. This is a single
integral relation which must be satisfied by profile
functions $\rho_{\pm}(\beta)$. The only non-zero components of
energy-momentum tensor are therefore:
\begin{eqnarray}
&&T^{tt}=r^{-2}C_{1},\ T^{rr}=r^{-2}C_{2},\label{Tres}
\end{eqnarray}
where the constants $C_{1,2}>0$ and
\begin{eqnarray}
&&C_{1}=\int d\beta(\rho_{+}(\beta)\cosh^{2}\beta
+\rho_{-}(\beta)\sinh^{2}\beta),\nn\\
&&C_{2}=\int d\beta(\rho_{+}(\beta)\sinh^{2}\beta
+\rho_{-}(\beta)\cosh^{2}\beta),\label{C120}\\
&&C_{0}=\int d\beta(\rho_{+}(\beta)+\rho_{-}(\beta))\sinh\beta\cosh\beta
=0.\nn
\end{eqnarray}
Note that condition of energetic balance $C_{0}=0$ can be satisfied
also when the inflow of normal matter is completely switched off:
$\rho_{+}(\beta)=0$. In this case the energetic balance must be satisfied 
by tachyonic flows: incoming flow of tachyons should have the same total 
energy as outgoing flow of tachyons. Further we give several examples 
of flow distributions satisfying all necessary conditions.

\vspace{2mm}\noindent
{\it Example 1:} tachyonic flow with symmetric profile
\begin{eqnarray}
&&\rho_{+}(\beta)=0,\ \rho_{-}(\beta)=\rho_{-}(-\beta).\nn
\end{eqnarray}

\vspace{2mm}\noindent
{\it Example 2:} flow of normal matter with symmetric profile
\begin{eqnarray}
&&\rho_{+}(\beta)=\rho_{+}(-\beta),\ \rho_{-}(\beta)=0.\nn
\end{eqnarray}
Note that this generally requires a presence of outflow for normal matter.
This scenario is only possible if the incoming flow of normal matter 
turns back before reaching event horizon.

\begin{figure}
\centering
\includegraphics[width=\textwidth]{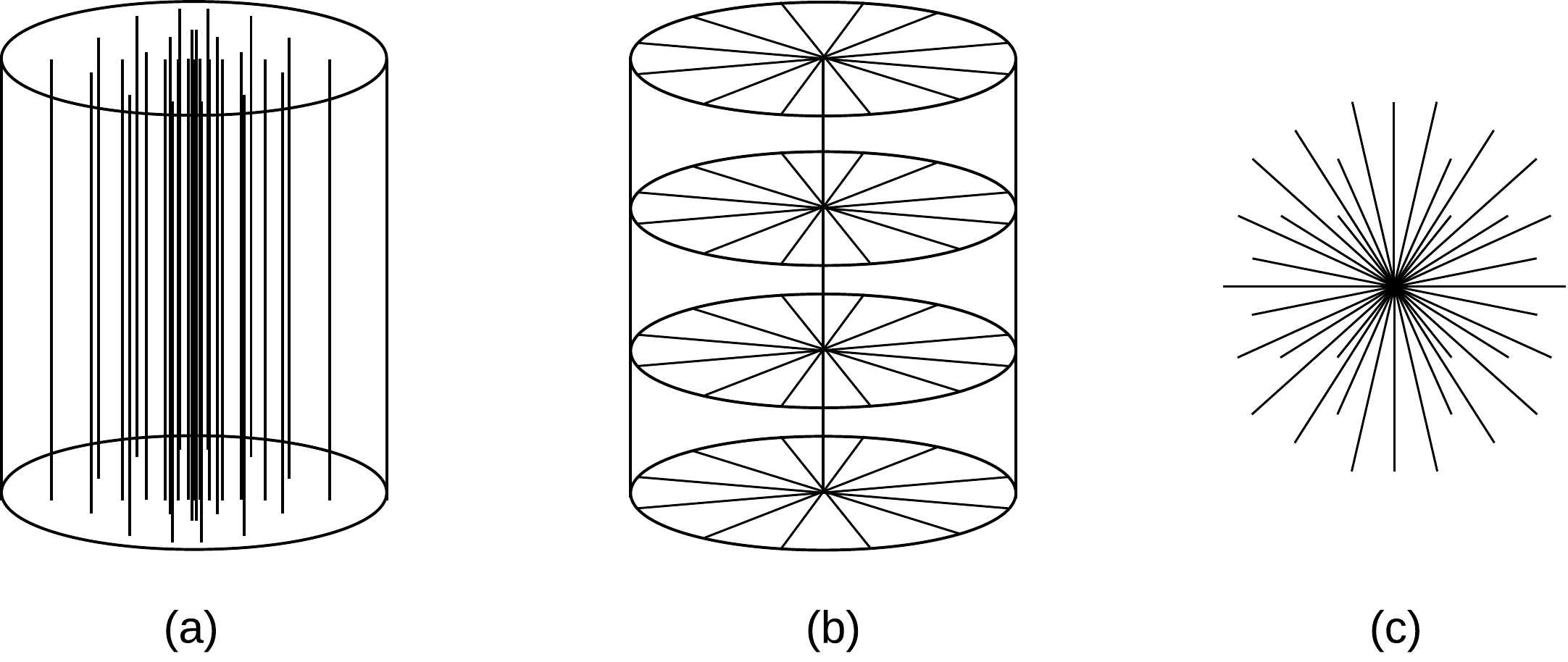}
\caption{Examples of flow distributions: (a) normal matter in slow limit, 
(b,c) ``tachyonic monopole''.}
\label{f2}
\end{figure}

\vspace{2mm}\noindent
{\it Example 3:} normal matter in slow limit 
\begin{eqnarray}
&&\rho_{+}(\beta)=\delta(\beta),\ \rho_{-}(\beta)=0.\nn
\end{eqnarray}
A marginal scenario, depicted on Fig.\ref{f2}a. The only possible flow
configuration without tachyons and without outgoing flow of normal matter.
Can be considered as incoming flow in the limit $\beta\to-0$.
Note that spatial distribution of matter here is not arbitrary
and must satisfy $\rho(r)\sim r^{-2}$. The case corresponds to 
$C_{1}=1$, $C_{2}=0$.

\vspace{2mm}\noindent
{\it Example 4:} ``tachyonic monopole''
\begin{eqnarray}
&&\rho_{+}(\beta)=0,\ \rho_{-}(\beta)=\delta(\beta).\nn
\end{eqnarray}
Scenario depicted on Fig.\ref{f2}b. 
The world lines go in purely spatial direction,
orthogonally to time axis. In spatial projection this configuration looks 
like a point surrounded by radially diverging tachyonic fibers, Fig.\ref{f2}c. 
This case corresponds to $C_{1}=0$, $C_{2}=1$.

\begin{figure}
\centering
\includegraphics[width=\textwidth]{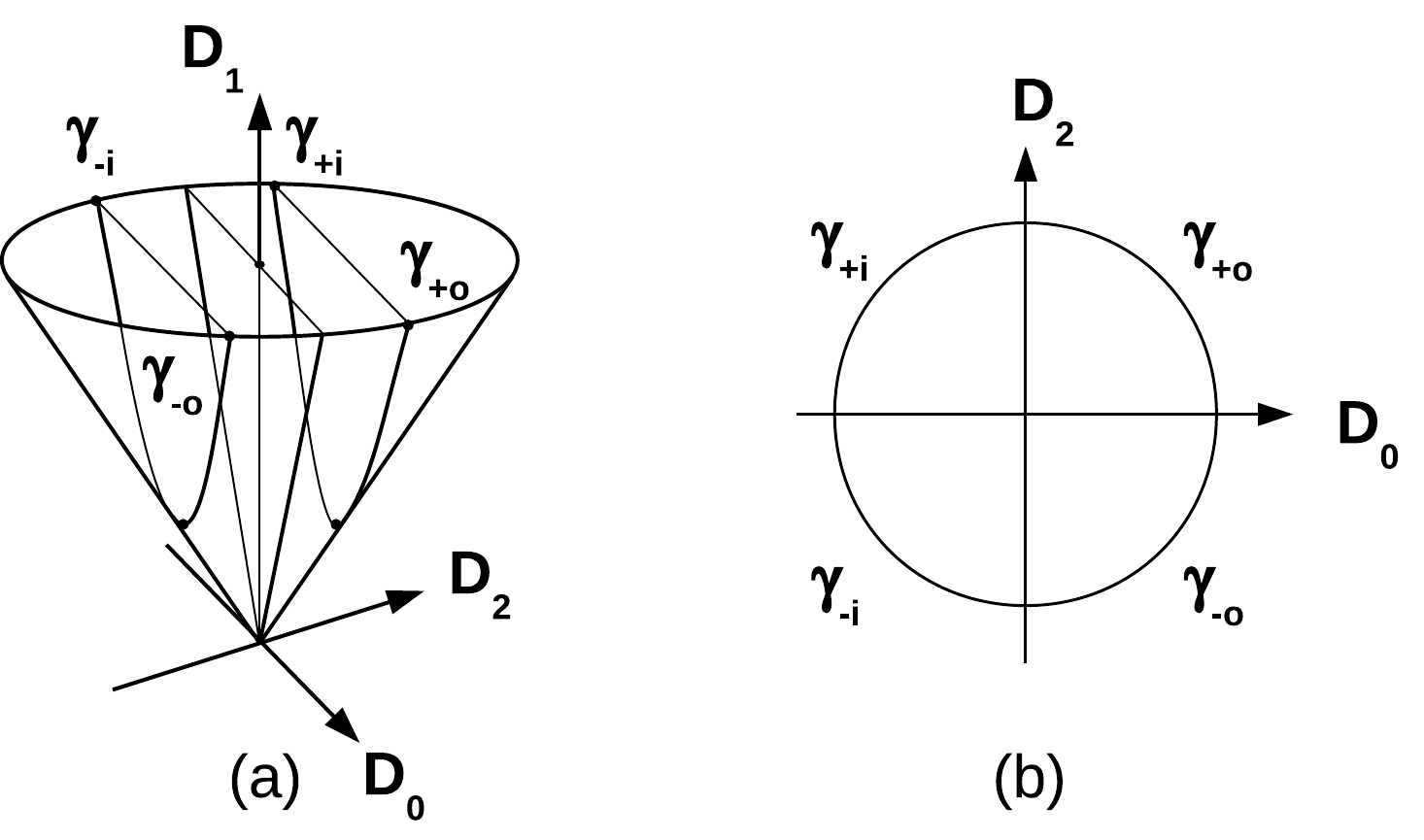}
\caption{Region of parameter variation: 
(a) in space $(D_{1},D_{2},D_{0})$,
(b) in cross-section $D_{1}=1$.
}
\label{f3}
\end{figure}

\begin{figure}
\centering
\includegraphics[width=\textwidth]{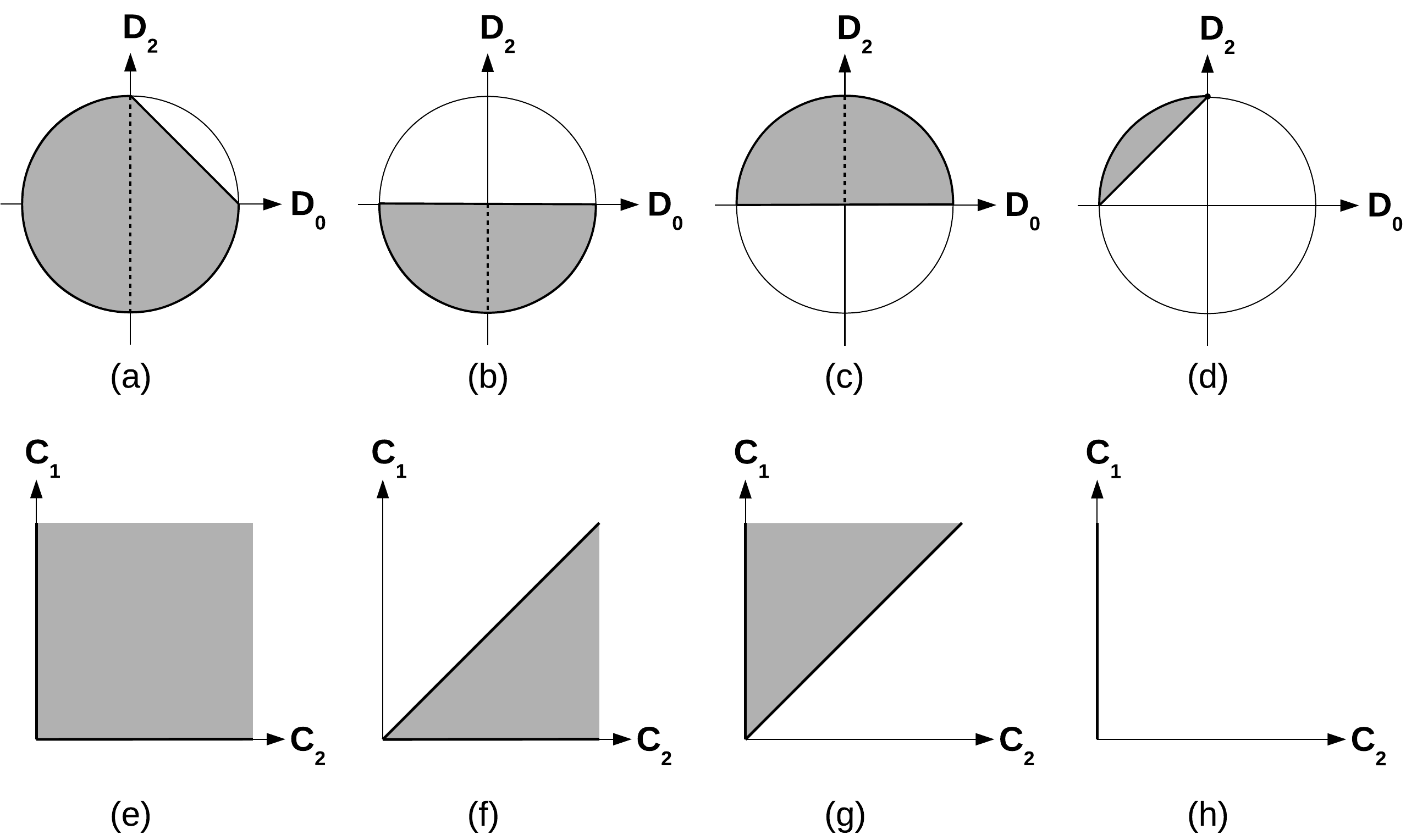}
\caption{Regions of parameter variation for various scenarios
(see description in text).}
\label{f4}
\end{figure}

\vspace{2mm}\noindent
Further we will clarify which regions on the plane $(C_{1},C_{2})$
can be occupied by different types of flow distributions.
Let's consider (\ref{C120}) as a mapping of non-negative functions
$\rho_{\pm}(\beta)$ to 3-dimensional space:
\begin{eqnarray}
&&(C_{1},C_{2},C_{0})
=\int d\beta(\rho_{+}(\beta)\gamma_{+}(\beta)
+\rho_{-}(\beta)\gamma_{-}(\beta)),\nn\\
&&\gamma_{+}(\beta)=(\cosh^{2}\beta,\sinh^{2}\beta,\sinh\beta\cosh\beta),\nn\\
&&\gamma_{-}(\beta)=(\sinh^{2}\beta,\cosh^{2}\beta,\sinh\beta\cosh\beta).\nn
\end{eqnarray}
Performing transformations
\begin{eqnarray}
&&D_{1}=C_{1}+C_{2},\ D_{2}=C_{1}-C_{2},\ D_{0}=2C_{0},\nn
\end{eqnarray}
we have
\begin{eqnarray}
&&(D_{1},D_{2},D_{0})
=\int d\beta(\rho_{+}(\beta)\gamma_{+}(\beta)
+\rho_{-}(\beta)\gamma_{-}(\beta)),\nn\\
&&\gamma_{\pm}(\beta)=(\cosh2\beta,\pm1,\sinh2\beta).\label{convhull}
\end{eqnarray}
The curves $\gamma_{\pm}$
form conic sections, shown on Fig.\ref{f3}a.
The figure also shows different segments of the curves,
corresponding to inflows and outflows: $\gamma_{\pm i,o}$.
Vectors $\rho_{\pm}\gamma_{\pm}$ with $\rho_{\pm}>0$ define
rays from the origin to the points of the curves.
The mapping (\ref{convhull}) defines a convex hull of these rays. 
The result will be different dependently on which parts of the curves 
are taken in the scenario.
The cone has an equation:
\begin{eqnarray}
&&D_{1}^{2}=D_{2}^{2}+D_{0}^{2}.\nn
\end{eqnarray}
In cross-section $D_{1}=1$ the problem is reduced to taking 
convex hulls of corresponding circular arcs, see Fig.\ref{f3}b. 
Further we need to take 
a cross-section $D_{0}=0$, representing the equation
of energetic balance. Transforming the result in original coordinates,
we obtain the regions on a plane $(C_{1},C_{2})$ we are looking for.
Several possibilities are considered on Fig.\ref{f4}.

\vspace{2mm}\noindent
{\it Case 1:} Fig.\ref{f4}a,e,
inflow of normal matter, inflow and outflow of tachyons
\begin{eqnarray}
&&C_{1}\geq0,\ C_{2}\geq0.\nn
\end{eqnarray}

\vspace{2mm}\noindent
{\it Case 2:} Fig.\ref{f4}b,f, only tachyons, inflow and outflow
\begin{eqnarray}
&&C_{2}\geq C_{1}\geq0.\nn
\end{eqnarray}

\vspace{2mm}\noindent
{\it Case 3:} Fig.\ref{f4}c,g, only normal matter, inflow and outflow
\begin{eqnarray}
&&C_{1}\geq C_{2}\geq0.\nn
\end{eqnarray}
This scenario is only possible if the incoming flow of normal matter 
turns back before reaching event horizon.

\vspace{2mm}\noindent
{\it Case 4:} Fig.\ref{f4}d,h, only normal matter, inflow,
the marginal case from Example~3
\begin{eqnarray}
&&C_{1}\geq0,\ C_{2}=0.\nn
\end{eqnarray}

\vspace{2mm}\noindent
The limiting lines on these plots correspond to
\begin{itemize}
\item $C_{1}\geq0,\ C_{2}=0$, slow normal matter
\item $C_{2}\geq0,\ C_{1}=0$, ``tachyonic monopole''
\item $C_{1}=C_{2}\geq0$, Cases 2,3,  
a limit of lightlight particles, $\beta\to\infty$.
\end{itemize}

\section{Solving Einstein field equations}\label{sec_efe}

The equations have a form:
\begin{eqnarray}
&&R_{\mu\nu}-\half g_{\mu\nu}R=8\pi G T_{\mu\nu},\label{efe1}
\end{eqnarray}
where $g_{\mu\nu}$ is metric tensor, $R_{\mu\nu}$ is Ricci curvature tensor,
$R=g^{\mu\nu}R_{\mu\nu}$, 
$T_{\mu\nu}$ is energy-momentum tensor, 
$G$ is gravitational constant. Further we fix a system of units 
$4\pi G=1$. Ricci tensor is a sophisticated non-linear 
function of metric tensor and its first and second derivatives, whose explicit
expression can be found in \cite{Dirac,MB}.

Before proceeding to solution, there are some introductory remarks.
At first, not all components of Einstein field equations are independent.
There is a compatibility requirement, equivalent to continuity condition
on energy-momentum tensor: $\nabla_{\mu}T^{\mu\nu}=0$. 
For the flows of free falling particles 
this condition is equivalent to the motion of particles along geodesics. 
Thus, the system (\ref{efe1}) incorporates a condition
on matter distribution, the flow must be geodesic.

Secondly, GR is invariant under diffeomorphisms of coordinates and
corresponding transformations of metric tensor. A set of general solutions
of Einstein field equations contains together with every solution
all its diffeomorphisms. To fix this freedom, gauge conditions are selected, 
equivalent to a choice of particular coordinate system,
e.g. synchronous coordinates $g_{0i}=0,\ i>0$.

In this paper we will solve not the general system (\ref{efe1}),
but its linearization. Namely, we will consider slightly curved
metric, represented in the form $g_{\mu\nu}=\eta_{\mu\nu}+h_{\mu\nu}$,
where $\eta_{\mu\nu}$ is the flat metric and $h_{\mu\nu}$ is a small correction. 
Energy-momentum tensor from the previous section
corresponds to geodesic flows of particles in flat metric. 
We substitute this matter contribution to the right hand side of (\ref{efe1}),
consider it as small correction to vacuum case 
and solve the system for the linear term 
$h_{\mu\nu}$. According to approximation schemes \cite{PN}, 
this solution can be used further to correct the geodesics and compute
higher order terms. In this paper we restrict ourselves to the investigation
of linear correction and its influence to the motion of probe particles.

For spherically symmetric stationary problems one can 
choose the metric in the form \cite{Dirac,MB}:
\begin{eqnarray}
&&ds^{2}=-A(r)dt^{2}+B(r)dr^{2}+r^{2}(d\theta^{2}+\sin^{2}\theta d\phi^{2}),\nn
\end{eqnarray}
where after substitution
\begin{eqnarray}
&&A(r)=e^{2h(r)}f(r),\ B(r)=f(r)^{-1},\ f(r)=1-2m(r)/r\nn
\end{eqnarray}
the system (\ref{efe1}) is reduced to
\begin{eqnarray}
&&m'(r)=r^{2}(-T^{t}_{t}),\ 
h'(r)=rf(r)^{-1}(-T^{t}_{t}+T^{r}_{r}).\nn
\end{eqnarray}
We substitute here the components of energy-momentum tensor 
(\ref{Tres}) and use flat metric to raise and lower 
the indices: $T^{t}_{t}=-T^{tt}$, $T^{r}_{r}=T^{rr}$. The difference
between real and flat metric being multiplied to small $T$-components
becomes a higher order term, which can be neglected 
in considered approximation. Thus we have
\begin{eqnarray}
&&m'(r)=C_{1},\
h'(r)={C_{1}+C_{2}\over r-2m(r)}.\nn
\end{eqnarray}
Solution has a form:
\begin{eqnarray}
&&m(r)=C_{1}r+C_{3},\ h(r)=\epsilon\log |r-r_{0}|
+C_{4},\nn
\end{eqnarray}
with two new integration constants $C_{3,4}$ and
\begin{eqnarray}
&&\epsilon={C_{1}+C_{2}\over 1-2C_{1}},\ 
r_{0}={2C_{3}\over 1-2C_{1}}.\nn
\end{eqnarray}
The term $C_{3}$ corresponds to 
an arbitrary mass, located at the origin or distributed 
spherically symmetrically under $r<r_{1}$, inside the inner sphere 
of the considered spherical layer.
The constant $C_{4}$ can be absorbed in definition of time
and fixed arbitrarily, e.g by requiring $h(r_{1})=0$. 

If, for a moment, we set the constants $C_{1,2}$ to zero, these formulae 
reconstruct well known Schwarzschild's solution for spherically symmetric 
black hole, with parameter $r_{0}$ representing Schwarzschild's radius:
\begin{eqnarray}
&&r_{0}=2C_{3},\ m(r)=C_{3},\ h(r)=0,\nn\\
&&A(r)=(1-r_{0}/r),\ B(r)=(1-r_{0}/r)^{-1}.\nn
\end{eqnarray}

Further, considering the case of non-zero $C_{1,2}$, we should keep
the solution in the frames of admitted approximation, 
the metric should deviate only slightly from the flat case. 
This implies $|A(r)-1|\ll1$, $|B(r)-1|\ll1$,
or equivalently $|2m(r)/r|\ll1$, $|h(r)|\ll1$ 
everywhere in the considered range $r\in[r_{1},r_{2}]$.
This can be achieved by fixing $r_{1}\ll r_{2}$ and selecting
sufficiently small constants satisfying
$C_{1}\ll1$, $2C_{3}\ll r_{1}$, $(C_{1}+C_{2})\log r_{2}/r_{1}\ll1$.
The first condition keeps us away from the pole appeared 
in $r_{0}$-definition, the second one requires that the considered 
spherical layer is well above Schwarzschild's radius
and the third one ensures that our matter distribution produces small curvature 
of space-time in between $r_{1}$ and $r_{2}$. Equivalently, one can select
$C_{1}\ll1$, $r_{0}\ll r_{1}$, $\epsilon\ll(\log r_{2}/r_{1})^{-1}$. 
In this limit we have
\begin{eqnarray}
&&\epsilon=C_{1}+C_{2},\ r_{0}=2C_{3},\ 
h(r)=\epsilon\log r/r_{1},\ f(r)=1-2C_{1}-r_{0}/r,\nn\\
&&A(r)=1-2C_{1}+2\epsilon\log r/r_{1}-r_{0}/r,\
B(r)=1+2C_{1}+r_{0}/r.\nn
\end{eqnarray}
Thus we have for temporal component of metric tensor
\begin{eqnarray}
&&g_{00}=-A(r)=-1+2C_{1}-2\epsilon\log r/r_{1}+r_{0}/r.\nn
\end{eqnarray}
This component is related with gravitational potential, describing
geodesic motion of non-relativistic probe particles \cite{Dirac,MB}: 
\begin{eqnarray}
&&g_{00}=-1-2\phi,\ \ddot x=-\mbox{grad}\;\phi,\nn
\end{eqnarray}
thus we have
\begin{eqnarray}
&&\phi=-C_{1}+\epsilon\log r/r_{1}-r_{0}/(2r),\nn
\end{eqnarray}
the probe particles possess acceleration
directed radially towards the origin
\begin{eqnarray}
&&a_{r}=\epsilon/r+r_{0}/(2r^{2}).\nn
\end{eqnarray}
Here the second term corresponds to Newton's law,
becoming $GM/r^{2}$ after reconstruction of physical units.
The first term is the effect of matter distribution constructed 
in our model. Considering circular orbits around the origin
and substituting $a_{r}=v^{2}/r$, we have for the orbital velocity
\begin{eqnarray}
&&v^{2}=\epsilon+r_{0}/(2r).\label{vres}
\end{eqnarray}
At large $r$ the velocity does not tend to zero,
as it should be for purely Newtonian case.
Instead, it tends to a positive constant value. 

\begin{figure}
\centering
\includegraphics[width=\textwidth]{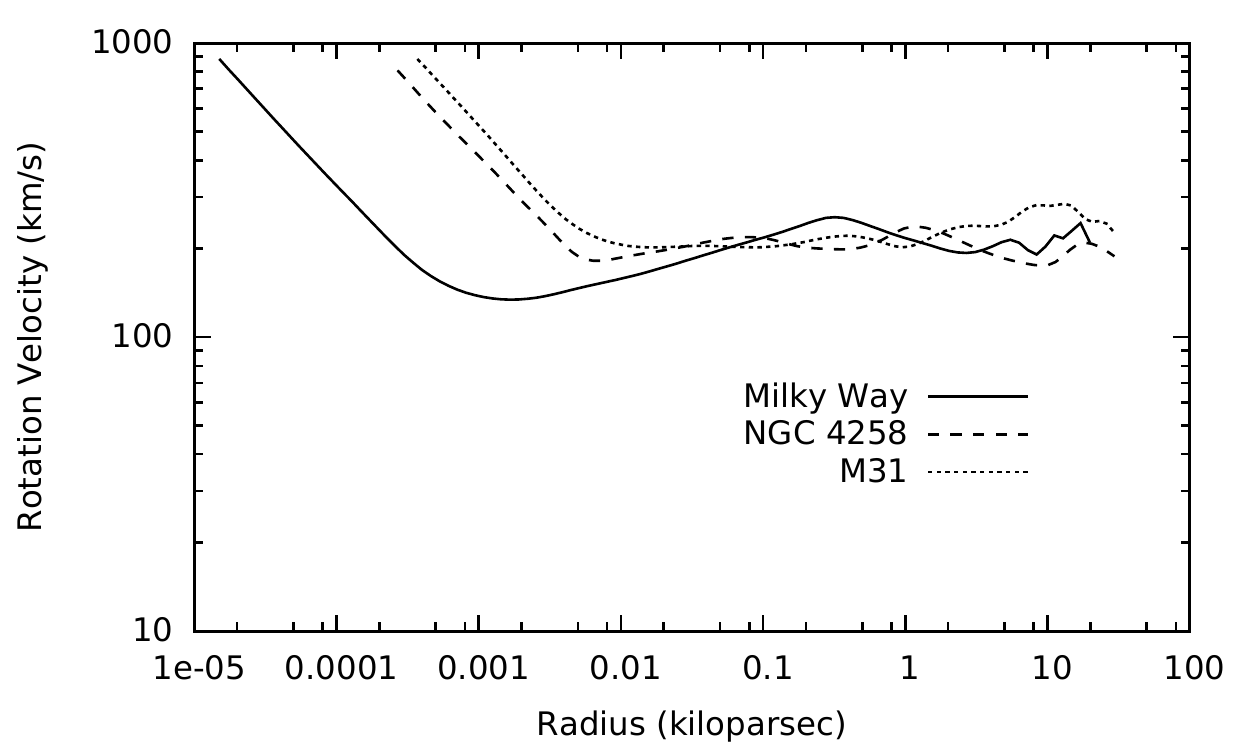}
\caption{Measured velocities of stars in galaxies, as a function of distance to the center, data from \cite{VR2}.}
\label{f5}
\end{figure}

\section{Discussion}\label{sec_disc}
Dependence of orbital velocity on radius with asymptotic transition 
to non-zero constant shows a similarity with
the measured rotation curves of the galaxies, see fig.\ref{f5}. 
In 1978 Vera Rubin and coworkers have shown that the velocities 
of stars and interstellar gas in high-luminosity
spiral galaxies are constant in wide range of distances \cite{VR3}.
The estimation involving only luminous matter provided much smaller 
velocities and the rotation curves falling with the distance. 
Attempts to explain this discrepancy gave birth to the concept of
hidden mass, also known as dark matter.
As we see, tachyonic models are well suited for the role of dark matter. 
Already our simple model describes important qualitative features
as increased orbital velocity and asymptotically constant rotation curve.
Of course, this model is still too
idealized for comparison with a real galaxy, in fact, from the necessary 
elements it contains only the central supermassive black hole.
To explain fine details of the rotation curves, one should take into account
the distribution of luminous matter, the presence of other 
black holes in the galaxy and the influence of gravitational field
to the shape of tachyonic world lines.

One such fine detail can be an observable deviation of rotation curve 
from the constant. According to \cite{VR2}, this deviation
depends on luminosity of the galaxy: most luminous galaxies
have slightly decreasing rotation curves, intermediate luminosities
correspond to constant rotation curves, low-luminosity galaxies have increasing 
rotation curves. In particular, low-luminosity dwarf galaxy M33 
shows slightly increasing rotation curve \cite{M33}. 
The measurement of 21 spiral galaxies of Sc type 
shows that most of them have slightly increasing rotation curves \cite{VR1}.
The deviation of rotation curve from the constant can be explained
by the presence of the other black holes, i.e. sources and sinks 
of dark matter distributed
over the galaxy, which can lead to the dark matter term $\epsilon(r)$ 
dependent on the distance to the center of the galaxy.
In weak field approximation the sources contribute additively 
to the gravitational potential and a sum of isotropic sources
will give the dark matter term increasing with the distance,
providing the increasing rotation curve.
On the other hand, if the tachyonic world lines sourced by the central
black hole will sink in the other black holes distributed over the galaxy, 
the distribution of dark matter can be truncated and one can see 
falling rotation curves outside of truncation radius. 
These scenarios will require more complex computations, based on 
non-isotropic flows and non-straight tachyonic world lines.

At a larger scale dark matter forms superstructures, they
look like a network of filaments connecting the galaxies \cite{filaments}. 
Such spacelike structures can be composed of tachyonic world lines stretched 
between the galactic black holes. Theoretically, these networks can also 
connect white holes and other places where conditions 
are hot enough, Big Bang, Big Crunch, etc. In the models describing
multiple universes \cite{multiverse} tachyonic world lines 
will not be confined in one universe and can pass from one universe to another.
Analysis of such scenarios would also require more sophisticated methods
and presumably can be done only with the aid of numerical simulations.

The calculations in our paper were done in the limit of weak fields and 
were similar to those in Newtonian limit. However, the matter distribution
involved superluminal particles and Newtonian limit was not applicable
as is. The work \cite{PN} mentions a combination of different approximations:
weak field, near zone, small $v/c$; here we used just the first option.
It is interesting to continue the model in the region of 
strong fields and to look what happens with tachyons under event horizon.

We remind that initial parameters of the model were distributions
$\rho_{\pm}(\beta)$. They were contracted to two constants in tensor 
of energy-momentum,
so that the metric actually depends only on two parameters $C_{1,2}$.
They were summed in gravitational potential to a single constant $\epsilon$.
All these parameters are free, they can be restricted only by inequalities,
described in Section \ref{sec_set}. The inequalities appear after {\it a priori}
restrictions on the structure of the flows, e.g. $C_{1}\leq C_{2}$, 
if the incoming flow of normal matter is completely switched off.
On the other hand, the constants can be fixed by a detailed model
describing the processes under event horizon 
(e.g. explaining a proportion between normal and dark components of the flow)
and also by external boundary conditions on incoming and outgoing flows
(e.g. connecting the flows from different black holes). 
In this way one can obtain
a picture of the universe as a global relativistic network,
a cosmic web of tachyonic filaments stretched between the black holes and 
the galaxies around them. 
Calculation of equilibrium of such network using analytical and 
numerical methods would be a challenging problem.

\section{Conclusion}\label{sec_concl}
We have considered a spherically symmetric stationary problem, including a black hole, incoming and outgoing flows of tachyons and optionally incoming flow of normal matter. Computations in the limit of weak field show that probe particles moving along circular orbits in this model have a dependence of orbital velocity on a distance identical with the typical rotation curves of galaxies. We have discussed a possibility to use the model for a description of dark matter distribution in galaxies and the extensions of the model for a description of more complex scenarios.

\end{document}